\documentclass[seceq]{ptptex}

\usepackage{graphicx}

\title{
Static Exact Solutions of a Spin Model Exhibiting Glassy Dynamics%
}

\author{
Chiaki \textsc{Yamaguchi}%
}

\inst{
Kosugichou 1-359, Kawasaki 211-0063, Japan
}

\abst{
A spin model which exhibits glassy dynamics has been
 proposed by Newman and Moore.
 This model possesses no randomness for exchange interactions.
 We propose partial trace methods for 
 obtaining static exact solutions of this model under free boundary conditions,
 and show some static exact solutions.
 We obtain that, from the exact solutions of
 the specific heat and the correlation functions,
 the transition temperature is zero, although
 the thermal average of a spin is zero when the temperature is zero.
 We investigate the ground states.
 In the ground states,
 a part of the present results especially disagrees with
 the part of the previous results. We discuss the disagreements.
}

\begin{document}
\maketitle

\section{Introduction}

Spin glass models\cite{EA, SK},
 that are known as popular models for glassy systems,
 have randomness for
 exchange interactions.
 The mathematical study is hard. 
 In addition, there is a problem whether
 the randomness for
 exchange interactions is necessity or not for glassy properties.
 Therefore,
 searching and investigating
 models, that have the properties of glassy systems
 and have no randomness for exchange interactions,
 can be one of promising methods for understandings of glassy systems.
 For achieving this purpose,
 a spin model which has no randomness for exchange interactions
 and exhibits glassy dynamics has been
 proposed by Newman and Moore\cite{NM}.
 For dynamical features,  
 this model has aging behavior \cite{NM}, and
 it falls out of equilibrium at a temperature which
 decreases logarithmically as a function of the cooling time \cite{NM}, 
 for example.  
 Dynamical features of this model
 are investigated in Refs.~\citen{NM, GN, G, JG}.

For static solutions, the solutions
 have been exactly obtained on
 lattices of length a power of two
 along at least one dimension
 under periodic boundary conditions \cite{NM, GN, G, JG}.
 In other words, this model has a special property.
 Therefore, this model has been solved in a special case
 that, under periodic boundary conditions,
 the lattice length is a power of two
 along at least one dimension.
 There have been no solutions on 
 lattices of any length under 
 periodic boundary conditions.
 In addition,
 there have been no solutions on 
 lattices of any length under 
 free boundary conditions.
 Since the previous method for solving this model
 uses a special property of 
 this model,
 there is a possibility that 
 a part of the results is different from that on
 lattices of any length, and
 this issue has not been shown exactly.
 In this article,
 by solving exactly this model on 
 lattices of any length under free boundary conditions,
 we exactly show that
 a part of the results is different from
 the part of the previous results.

We propose partial trace methods for 
 obtaining static exact solutions of this model under free boundary conditions, 
 and find some static exact solutions.
 We find the exact solutions under free boundary conditions for
 the energy, the specific heat,
 the free energy, the entropy,
 the number of ground-state degeneracy,
 the thermal average of a spin,
 the two-point correlation function,
 the three-point correlation function
 and the four-point correlation function.
 We examine the transition temperature, and investigate the ground states.
 We discuss the disagreements between a part of
 the present results and
 the part of the previous results.

This article is organized as follows.
 First in \S\ref{sec:2}, the model is explained.
 In \S\ref{sec:3}, we describe the present partial trace methods for solving this model
 and obtain results.
 In \S\ref{sec:4}, the disagreements between a part of the present results
 and the part of the previous results are discussed. 
 This article is summarized in \S\ref{sec:5}.

\section{Model} \label{sec:2}

The Hamiltonian of this model under free boundary conditions, ${\cal H}$, 
 is given by
\begin{equation}
 {\cal H} = J
 \sum^{L - 1}_{x = 1} \sum^{L - 1}_{y = 1} 
 \sigma_{x, y} \, \sigma_{x + 1, y} \, \sigma_{x, y + 1} \, ,
 \label{eq:Hamiltonian}
\end{equation}
 where $\sigma_{x, y}$ is an Ising spin,
 $\sigma_{x, y} = \pm 1$, and $x$ and $y$ are
 coordinates of the spin. We set $J > 0$.
 This system is on downward pointing-triangles of
 the triangular lattice.
 $L$ is the linear size of this system.
 The number of spins, $N$, is $L^2 - 1$,
 and the number of plackets, $N_P$, is $(L-1)^2$.

\begin{figure}[t]
\begin{center}
\includegraphics[width=0.50\linewidth]{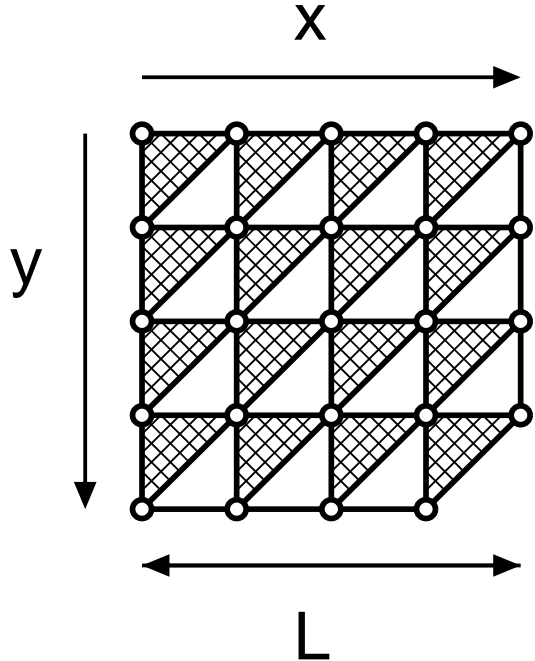}
\end{center}
\caption{
The present model.
 In this figure,
 the linear size of this system, $L$,
 is $5$, the number of spins, $N$, is $24$,
 and the number of plackets, $N_P$, is $16$.
 The plackets are drawn as meshes.
 \label{fig:zu1}
}
\end{figure}

Figure~\ref{fig:zu1} shows the present model\cite{Chuusyaku1}.
 In this figure,
 the linear size of this system, $L$,
 is $5$, the number of spins, $N$, is $24$,
 and the number of plackets, $N_P$, is $16$.
 The plackets are drawn as meshes.

Note that this model is distinct from the model of
 Baxter and Wu\cite{BW}, which has interactions on upward-pointing triangles
 also. Its spin behavior of the model of
 Baxter and Wu is entirely different from this model \cite{NM}.

In the ground state,
 there is a relation among spins given by \cite{NM}
\begin{equation}
 \sigma_{x, y}  \sigma_{x + 1, y} \sigma_{x, y + 1} = - 1 \, .
 \label{eq:groundspins}
\end{equation}
 From this relation,
 if the boundary spins of system are given,
 all the spins in the ground state are uniquely obtained.

This model displays glassy behavior under single-spin-flip dynamics.
 It has been pointed out that, in Ref.~\citen{NM}, 
 this model has aging behavior, and
 it falls out of equilibrium at a temperature which
 decreases logarithmically as a function of the cooling time.  
 See Refs.~\citen{NM, GN, G, JG} for dynamical features of this model.

The transformed models by a transformation $J \to \pm J$ for 
 each three-spin interaction have identical behavior with this model\cite{NM}.

This model under a magnetic field has been also proposed
 concerned with thermodynamic transitions associated with
 irregularly ordered ground states\cite{S}.

\section{Results} \label{sec:3}

We describe the present partial trace methods for solving this model
 and obtain results.

We adopt the free boundary conditions as the boundary conditions.
 Therefore, there is a spin which is only on a placket.
 We perform the integration of the spin at first.
 We define the spin as $\sigma_i$, and define a function
 $A (\sigma_j, \sigma_k, \{ \sigma_l \})$, where
 $\{ \sigma_l \}$ is the set of all spins except
 for $\sigma_i$, $\sigma_j$ and $\sigma_k$.
 We introduce a variable $K$ and define $K \equiv \beta J$,
 where $\beta$ is the inverse temperature,
 and $\beta = 1 / k_B T$.
 $k_B$ is the Boltzmann constant, and $T$ is the temperature.
 Then, we perform the integration of the spin $\sigma_i$ as
\begin{eqnarray}
 & & \sum_{\sigma_i} \sum_{\sigma_j} \sum_{\sigma_k}
 e^{- K \sigma_i \sigma_j \sigma_k}
 A  (\sigma_j, \sigma_k, \{ \sigma_l \}) \nonumber \\
 &=& \sum_{\sigma_j} \sum_{\sigma_k}
 (e^{- K \sigma_j \sigma_k} + e^{K \sigma_j \sigma_k})
 A  (\sigma_j, \sigma_k, \{ \sigma_l \}) \nonumber \\
 &=& (e^{- K} + e^{K}) \sum_{\sigma_j} \sum_{\sigma_k}
 A  (\sigma_j, \sigma_k, \{ \sigma_l \}) \, . \label{eq:integ-1}
\end{eqnarray}
 In Eq.~(\ref{eq:integ-1}), we are able to treat
 $\sigma_j$ and $\sigma_k$ as the spins on the free boundaries.
 Therefore, we are also able to perform the integrations of $\sigma_j$ and $\sigma_k$
 by using similar integrations with the integration of $\sigma_i$
 after the integration of $\sigma_i$ is performed.

\begin{figure}[t]
\begin{center}
\includegraphics[width=0.75\linewidth]{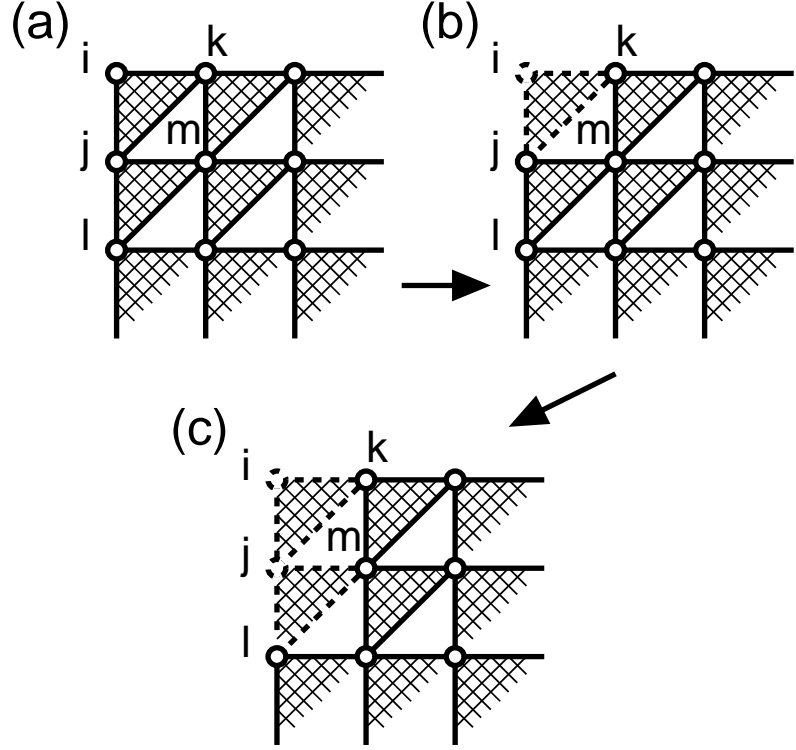}
\end{center}
\caption{
The partial trace method for Eq.~(\ref{eq:integ-1}).
(a) the figure before the integrations are performed. 
(b) the integration of the spin $i$ is performed. 
(c) the integrations of the spins $i$ and $j$ are performed. 
 \label{fig:zu2}
}
\end{figure}

Figure~\ref{fig:zu2} shows the partial trace method for Eq.~(\ref{eq:integ-1}). 
Figure~\ref{fig:zu2}(a) shows the figure before the integrations are performed. 
Figure~\ref{fig:zu2}(b) shows that the integration of the spin $i$ is performed.
Figure~\ref{fig:zu2}(c) shows that the integrations of the spins $i$ and $j$ are performed. 

By this partial trace method
 for Eq.~(\ref{eq:integ-1}),
 we are able to perform the integration of the partition function $Z$. 
 We obtain
\begin{eqnarray}
 Z &=& \sum_{ \{ \sigma_{x, y} \} }
 e^{- K \sum^{L - 1}_{x = 1} \sum^{L - 1}_{y = 1} 
 \sigma_{x, y} \sigma_{x + 1, y} \sigma_{x, y + 1} }
 \nonumber \\
 &=& 2^{N - N_P} (e^K + e^{- K})^{N_P} \, . \label{eq:Z_r1} 
\end{eqnarray}
 By using Eq.~(\ref{eq:Z_r1}), we obtain
\begin{equation}
 Z = 2^{2 (L - 1)} (e^K + e^{- K})^{(L - 1)^2} \, . \label{eq:Z_r2}
\end{equation}
By using Eq.~(\ref{eq:Z_r2}), we obtain the energy $E$ as
\begin{equation}
 E = - \frac{\partial}{\partial \beta} \ln Z
 = - (L - 1)^2 \, J \, {\rm tanh} K \, . \label{eq:E}
\end{equation}
 The energy per site in the thermodynamic limit is obtained as 
 $- J \, {\rm tanh} K$.
 This result agrees with the previous result \cite{NM}
 except for a constant and the coefficient of $J$. 
By using Eq.~(\ref{eq:E}), we obtain the specific heat $C$ as
\begin{equation}
 C = k_B \beta^2 \frac{\partial^2}{\partial \beta^2} \ln Z
 = k_B K^2 (L - 1)^2 \, {\rm sech}^2 K \, . \label{eq:C}
\end{equation}
By using Eq.~(\ref{eq:Z_r2}), we obtain the free energy $F$ as
\begin{equation}
 F = - \frac{1}{\beta} \ln Z
 = - \frac{J}{K} [ 2 (L - 1) \, \ln 2 +
 (L - 1)^2 \, \ln (e^K + e^{- K} ) ] \, . \label{eq:F}
\end{equation}
By using Eqs.~(\ref{eq:E}) and (\ref{eq:F}),
 we obtain the entropy $S$ as
\begin{eqnarray}
 S &=& \frac{k_B K}{J} ( E - F ) 
 \nonumber \\
 &=& 2 (L - 1) k_B \ln 2 + k_B (L - 1)^2
 \, [ \ln ( e^K + e^{- K} ) - K \tanh K ] \, .
\end{eqnarray}
 We define the ground-state entropy per site in the thermodynamic limit as $s^{(g)}$.
 We obtain
\begin{equation}
 s^{(g)} = \lim_{K \to \infty} \lim_{N \to \infty} \frac{S}{N} = 0 \, .
\end{equation}
 The number of ground-state degeneracy, $\Omega$, is obtained as
\begin{equation}
 \Omega = \lim_{K \to \infty} e^{S / k_B} = 2^{2 (L -1)} \, .
 \label{eq:Omega}
\end{equation}
 Therefore, the number of ground-state degeneracy is not one,
 although $s^{(g)} = 0$.

We calculate the thermal average of a spin, $\langle \sigma_{x, y} \rangle_T$.
At first, we perform the integrations of all the spins except for the focused spin
 $\sigma_{x, y}$ by
 applying the partial trace method for Eq.~(\ref{eq:integ-1}).
 Therefore, we obtain 
\begin{equation}
 \langle \sigma_{x, y} \rangle_T = 0 \, .
\end{equation}
 The present result shows $\langle \sigma_{x, y} \rangle_T = 0$ even when $T = 0$.

We calculate the two-point correlation function
 $\langle \sigma_{x, y} \sigma_{x', y'} \rangle_T$.
At first, we perform the integrations of all the spins except for the focused spins
 $\sigma_{x, y}$ and $\sigma_{x', y'}$
 by applying the partial trace method for Eq.~(\ref{eq:integ-1}).
 Therefore, we obtain 
\begin{equation}
 \langle \sigma_{x, y} \sigma_{x', y'} \rangle_T = \delta_{x, x'} \delta_{y, y'} \, ,
 \label{eq:TwoPCF}
\end{equation}
 where $\delta_{x, x'}$ is the Kronecker delta.

When calculations of more than three-spin correlations,
 we perform the integration of the spin $\sigma_i$, which is 
 one of the focused spins for thermal average, as
\begin{eqnarray}
 & & \sum_{\sigma_i} \sum_{\sigma_j} \sum_{\sigma_k}
 \sigma_i e^{- K \sigma_i \sigma_j \sigma_k}
 A  (\sigma_j, \sigma_k, \{ \sigma_l \}) \nonumber \\
 &=& \sum_{\sigma_j} \sum_{\sigma_k}
 (e^{- K \sigma_j \sigma_k} - e^{K \sigma_j \sigma_k})
 A  (\sigma_j, \sigma_k, \{ \sigma_l \}) \nonumber \\
 &=& (e^{- K} - e^{K}) \sum_{\sigma_j} \sum_{\sigma_k}
 \sigma_j \sigma_k A  (\sigma_j, \sigma_k, \{ \sigma_l \}) \, . \label{eq:integ-2}
\end{eqnarray}
 If $A  (\sigma_j, \sigma_k, \{ \sigma_l \})$ is
 a Boltzmann factor for $\sigma_j$, $\sigma_k$ and $\{ \sigma_l \}$,
 we are also able to perform
 the integrations of $\sigma_j$ and $\sigma_k$
 by using similar integrations with
 the integration of $\sigma_i$
 after the integration of $\sigma_i$ is performed.
 The partial trace methods for
 Eqs.~(\ref{eq:integ-1}) and (\ref{eq:integ-2}) are
 the partial trace methods that we propose in this article.

By applying the partial trace methods
 for Eqs.~(\ref{eq:integ-1}) and (\ref{eq:integ-2}), 
 we obtain 
\begin{equation}
  \langle \sigma_{x, y} \sigma_{x + 1, y} \sigma_{x, y + 1}
 \rangle_T = - \tanh K \, .
\end{equation}
 Similarly, we obtain
\begin{eqnarray}
 & & \langle \sigma_{x, y} \sigma_{x + 1, y} \sigma_{x, y + 1}
 \sigma_{x', y'} \sigma_{x' + 1, y'} \sigma_{x', y' + 1}
 \rangle_T \nonumber \\
 &=& ( \tanh K)^2 + [1 - ( \tanh K)^2]  \delta_{x, x'} \delta_{y, y'} \, .
\end{eqnarray}

We calculate the three-point correlation function
 and the four-point correlation function.
 The three-point correlation function has been already
 derived by a different method \cite{GN},
 and the four-point correlation function has not been derived.
 We are able to consider 
 the inside surrounded by the focused spins for thermal average 
 and the outside of the spins separately.
 By using two functions
 $A_{\rm outside}$ and $A_{\rm inside}$, we define
\begin{equation}
 A_{\rm outside} A_{\rm inside} \equiv 
 \langle \prod_{\{ (i, j) \}} \sigma_{i, j} \rangle_T Z \, ,
 \label{eq:AAZ}
\end{equation}
 where $\{ (i, j) \}$ is the set of coordinates
 of the focused spins for thermal average.
 $A_{\rm outside}$ is a weight for outside of the focused spins,
 and $A_{\rm inside}$ is a weight for inside surrounded by the focused spins.
 We define the number of spins for inside surrounded by the focused spins as $n$, and 
 define the number of plackets for inside surrounded by the focused spins as $n_P$.
 By applying the partial trace method for Eq.~(\ref{eq:integ-1}), we obtain
\begin{equation}
 A_{\rm outside} =
 2^{N - N_P - n + n_P} (e^K + e^{- K})^{N_P - n_P} \, . \label{eq:A_outside}
\end{equation}
 We define the number of applications of the partial trace method of Eq.~(\ref{eq:integ-1})
 for inside surrounded by the focused spins as $n_1$, and define
 the number of applications of the partial trace method of Eq.~(\ref{eq:integ-2})
 for inside surrounded by the focused spins as $n_2$.
 There is a relation, written as
\begin{equation}
 n_P = n_1 + n_2 \, . \label{eq:n_P}
\end{equation}
 If there is a correlation for the focused spins,
 we obtain $A_{\rm inside}$ as 
\begin{equation}
 A_{\rm inside} = 
 2^{n - n_1 - n_2} (e^K + e^{- K})^{n_1} (e^{- K} - e^K)^{n_2} \, .
 \label{eq:A_inside}
\end{equation}
 If there is no correlation for the focused spins,
 we obtain $A_{\rm inside}$ as
\begin{equation}
 A_{\rm inside} = 0 \, . 
\end{equation}

At first, we calculate $A_{\rm outside}$.
 The calculation of $A_{\rm outside}$ is
 started from one of the spins on the free boundaries.
 $A_{\rm outside}$ is calculated
 by applying the partial trace method for Eq.~(\ref{eq:integ-1}).
 Next, we calculate $A_{\rm inside}$.
 The calculation of $A_{\rm inside}$
 is started from one of the focused spins for thermal average.
 $A_{\rm inside}$ is calculated
 by applying the partial trace methods for
 Eqs.~(\ref{eq:integ-1}) and (\ref{eq:integ-2}). 
 In a certain case, the partial trace method for Eq.~(\ref{eq:integ-1})
 is not applied to the calculation of $A_{\rm inside}$, while
 the partial trace method for Eq.~(\ref{eq:integ-2})
 is applied to the calculation of $A_{\rm inside}$.
 We find the spin correlations by examining the equations
 after the partial trace method for Eq.~(\ref{eq:integ-2}) is applied.

For the three-point correlation function,
 we generally obtain
\begin{equation}
 \langle \sigma_{x, y} \sigma_{x', y'} \sigma_{x'', y''}
 \rangle_T = 0 \, . \label{eq:ThreePCF1}
\end{equation}
 On the other hand, when $(x, y)$, $(x', y')$ and $(x'', y'')$ have 
 a special relation, there is a correlation.
 When coordinates of three spins are 
 $(x, y)$, $(x + 2^k, y)$, $(x, y + 2^k)$ and $k = 0, 1, 2, \ldots$,
 there are correlations.
 Then, by applying the partial trace methods for
 Eqs.~(\ref{eq:integ-1}) and (\ref{eq:integ-2}),
 we obtain $n = (2^{k - 1} + 1) (2^k + 1)$,
 $n_1 = 2^{k - 1} (2^k + 1) - 3^k$ and $n_2 = 3^k$.
 By using Eqs.~(\ref{eq:Z_r1}), (\ref{eq:AAZ}) - (\ref{eq:A_inside}),
 we obtain
\begin{equation}
 \langle \sigma_{x, y} \sigma_{x + 2^k, y} \sigma_{x, y + 2^k}
 \rangle_T = - ( \tanh K)^{3^k} \, , \quad k = 0, 1, 2, \ldots \, . 
 \label{eq:ThreePCF2}
\end{equation}
 This result agrees with the previous result obtained in Ref.~\citen{GN}.

For the four-point correlation function,
 we generally obtain
\begin{equation}
 \langle \sigma_{x, y} \sigma_{x', y'} \sigma_{x'', y''} \sigma_{x''', y'''}
 \rangle_T = 0 \, . \label{eq:FourPCF1}
\end{equation}
 On the other hand, when $(x, y)$, $(x', y')$, $(x'', y'')$
 and $(x''', y''')$ have 
 a special relation, there is a correlation.
 When coordinates of four spins are 
 $(x, y)$, $(x + 3 \cdot 2^k, y)$,
 $(x, y + 3 \cdot 2^k)$, $(x + 2^k, y + 2^k)$
 and $k = 0, 1, 2, \ldots$,
 there are correlations.
 Then, by applying the partial trace methods for
 Eqs.~(\ref{eq:integ-1}) and (\ref{eq:integ-2}), we obtain
 $n = (3 \cdot 2^{k - 1} + 1) (3 \cdot 2^k + 1)$,
 $n_1 = 3 \cdot 2^{k - 1} (3 \cdot 2^k + 1) - 6 \cdot 3^k$
 and $n_2 = 6 \cdot 3^k$.
 By using Eqs.~(\ref{eq:Z_r1}), (\ref{eq:AAZ}) - (\ref{eq:A_inside}),
 we obtain
\begin{equation}
 \langle \sigma_{x, y} \sigma_{x + 3 \cdot 2^{k}, y}
 \sigma_{x, y + 3 \cdot 2^{k}}
 \sigma_{x + 2^{k}, y + 2^{k}}
 \rangle_T = ( \tanh K)^{6 \cdot 3^k}  \, , \quad k = 0, 1, 2, \ldots \, .
 \label{eq:FourPCF2}
\end{equation}
 For the four-spin correlations,
 there are positive correlations.
 These positive correlations for spins have not been mentioned
 in the previous articles.

 From the solutions of
 the specific heat and the correlation functions,
 i.e., from Eqs.~(\ref{eq:C}), (\ref{eq:TwoPCF}),
 (\ref{eq:ThreePCF1}) - (\ref{eq:FourPCF2}),
 the transition temperature of this model is zero,
 although $\langle \sigma_{x, y} \rangle_T = 0$ even when $T = 0$.
 This result, that the transition temperature is zero,
 is concluded from 
 the thermal dependence of $\langle \sigma_{x, y} \rangle_T$ in Ref.~\citen{GN}.
 Note that, in Ref.~\citen{GN},
 $\langle \sigma_{x, y} \rangle_T = - 1$ when $T = 0$
 and $\langle \sigma_{x, y} \rangle_T = 0$ when $T > 0$
 are obtained.

 We investigate the spin configuration on a placket.
 We define the probability that 
 the spin configuration $(\sigma_{x, y}, \sigma_{x + 1, y}, \sigma_{x, y + 1})$
 on a placket is $( -1, -1, -1) $ as 
 $P_{x, y} ( -1, -1, -1)$. $P_{x, y} ( -1, -1, -1)$ is given by
\begin{equation}
 P_{x, y} ( -1, -1, -1) 
 = \frac{1}{8} (1 - \sigma_{x, y} ) (1 - \sigma_{x + 1, y} ) (1 - \sigma_{x, y + 1} ) \, .
 \label{eq:P-1-1-1}
\end{equation}
 We obtain
\begin{equation}
 \langle P_{x, y} ( -1, -1, -1) \rangle_T = \frac{e^{2 K}}{4 (e^{2 K } + 1)} \, .
  \label{eq:P-1-1-1T}
\end{equation}
 We define the probability that 
 the spin configuration $(\sigma_{x, y}, \sigma_{x + 1, y}, \sigma_{x, y + 1})$
 on a placket is $( -1, 1, 1) $ as 
 $P_{x, y} ( -1, 1, 1)$. $P_{x, y} ( -1, 1, 1)$ is given by
\begin{equation}
 P_{x, y} ( -1, 1, 1)
 = \frac{1}{8} (1 - \sigma_{x, y} ) (1 + \sigma_{x + 1, y} ) (1 + \sigma_{x, y + 1} ) \, .
 \label{eq:P-1+1+1}
\end{equation}
 We obtain
\begin{equation}
 \langle P_{x, y} ( -1, 1, 1) \rangle_T = \frac{e^{2 K}}{4 (e^{2 K } + 1)} \, .
 \label{eq:P-1+1+1T}
\end{equation}
 Similarly, we obtain
\begin{equation}
 \langle P_{x, y} ( 1, -1, 1) \rangle_T = 
 \langle P_{x, y} ( 1, 1, -1) \rangle_T =
 \frac{e^{2 K}}{4 (e^{2 K } + 1)} \, .
 \label{eq:P+1-1+1Tetc}
\end{equation}
 When $K \to \infty$, $\langle P_{x, y} \rangle_T$
 becomes the probability in the ground states.
 Therefore, in the ground states,
 the spin configurations 
 $(-1, -1, -1)$, $(-1, 1, 1)$, $(1, -1, 1)$ and $(1, 1, -1)$
 emerge with each 1-in-4 chance.

\section{Discussion} \label{sec:4}

There are disagreements between a part of the present results and
 the part of the previous results.
 We discuss the disagreements here.

The present result shows that
 the number of ground-state degeneracy is $2^{2 (L -1)}$.
 On the other hand, in Ref.~\citen{NM},
 it has been pointed out that the number of ground-state degeneracy
 is one.

The present result shows $\langle \sigma \rangle_T = 0$
 even when $T = 0$.
 In addition, the probability that the spin configuration
 $(\sigma_{x, y}, \sigma_{x + 1, y}, \sigma_{x, y + 1})$
 on a placket is $( -1, -1, -1) $
 is $1 / 4$ when $T = 0$.
 On the other hand, in Refs.~\citen{NM, GN},
 it has been pointed out that $\langle \sigma \rangle_T = -1$
 when $T = 0$.

This model has a special property.
 This model has been exactly solved in a special case that,
 under periodic boundary conditions,
 the lattice length is a power of two
 along at least one dimension, and has not been exactly solved
 on lattices of any length,
 so that
 there is a possibility that
 a part of the results on lattices of any length
 is different from
 the part of the previous results.
 From the results on lattices of any length
 under free boundary conditions,
 we see that there are the differences.
 Especially, the results for the ground states are different from 
 the previous results as described above.

 These disagreements between
 the present results and the previous results
 would not seriously affect the dynamics.
 The dynamics are decided by the energy changes of spin-flips
 \cite{NM}, and the energy changes of spin-flips are not mentioned in this
 article, but
 the energy changes of spin-flips are the same with the previous results.
 The number of ground-state degeneracy is 
 different from that of the previous result, but
 the ground state entropy in the thermodynamic limit
 is the same with the previous result, so that
 the disagreement for ground-state degeneracy
 would not seriously affect the dynamics. 
 Since the dynamics are decided by the energy changes of spin-flips,
 the effect from the different boundary conditions
 would not seriously affect the dynamics
 if the system size is large enough.

\section{Summary} \label{sec:5}

A spin model, which exhibits glassy dynamics
 and has no randomness for exchange interactions, has
 been proposed by Newman and Moore.
 We proposed partial trace methods for 
 obtaining static exact solutions of this model under free boundary conditions,
 and showed some static exact solutions.
 The partial trace methods for
 Eqs.~(\ref{eq:integ-1}) and (\ref{eq:integ-2})
 are the partial trace methods that we proposed in this article.
 We obtained that, from the exact solutions of
 the specific heat and the correlation functions,
 the transition temperature is zero, although
 the thermal average of a spin is zero when the temperature is zero.
 We investigated the ground states.
 In the ground states, a part of the present results especially disagreed with
 the part of the previous results.
 We discussed the disagreements, and
 concluded that 
 the disagreements would not seriously affect the dynamics.


\begin{thebibliography}{99}
\bibitem{NM}
 M. E. J. Newman and C. Moore,
 Phys.\ Rev.\ E \textbf{60} (1999), 5068.  
\bibitem{EA}
 S.~F.~Edwards and P.~W.~Anderson, J.\ of Phys.\ F \textbf{5} (1975), 965.
\bibitem{SK}
 D.~Sherrington and S.~Kirkpatrick, Phys.\ Rev.\ Lett.\ \textbf{35}
 (1975), 1792.
\bibitem{GN}
 J. P. Garrahan and M.E.J. Newman,
 Phys.\ Rev.\ E \textbf{62} (2000), 7670.
\bibitem{G}
 J. P. Garrahan, J.\ Phys.\ Condens.\ Matter\ \textbf{14} (2002), 1571.
\bibitem{JG}
 R. L. Jack and J. P. Garrahan,
 J.\ Chem.\ Phys.\ \textbf{123} (2005), 164508.
\bibitem{Chuusyaku1}
 Because, in Ref.~\citen{GN},
 spin correlations in this model are compared
 with the Pascal's triangle and
 it is convenient for us to investigate this model,
 y-axis is drawn as a downward pointing in this figure.
\bibitem{BW}
 R. J. Baxter and F. Y. Wu,
 Phys.\ Rev.\ Lett.\ \textbf{31} (1973), 1294.
\bibitem{S}
 S. Sasa, J.\ of Phys.\ A \textbf{43} (2010), 465002.
\end{thebibliography}
\end{document}